\documentclass[]{iopart}

\expandafter\let\csname equation*\endcsname\relax
  \expandafter\let\csname endequation*\endcsname\relax
\usepackage{amsmath}
\usepackage[english]{babel}
\usepackage[utf8]{inputenc}
\usepackage{graphicx}
\usepackage{epstopdf}
\usepackage{amssymb}
\usepackage{amsfonts}
\usepackage[mathcal]{eucal}
\usepackage{cite}
\usepackage[]{hyperref}
\usepackage{txfonts}
\usepackage{bm}
\allowdisplaybreaks[2]
\begin{document}

\newcommand{\bt}{\textbf}
\newcommand{\ep}{\varepsilon}
\newcommand{\vp}{\varphi}
\newcommand{\te}{\theta}
\newcommand{\vk}{\varkappa}
\newcommand{\si}{\sigma}
\newcommand{\om}{\omega}
\newcommand{\ra}{\rangle}
\newcommand{\la}{\langle}
\newcommand{\cd}{\!\cdot\!}
\newcommand{\op}[1]{\boldsymbol{\mathfrak{#1}}}
\newcommand{\fp}[2]{(#1\cdot#2)}
\newcommand{\thp}[2]{\bi #1\cdot\bi #2}
\newcommand{\opA}[1]{\boldsymbol{\mathsf{#1}}}
\newcommand{\opa}[1]{\mathsf{#1}}

\paper[Justification of the single-mode approximation]{Justification of the single-mode approximation for a finite-duration laser pulse interacting with an electron}

\author{O. D. Skoromnik$^1$ and I. D. Feranchuk$^2$}
\address{$^1$ Max Planck Institute for Nuclear Physics, Saupfercheckweg 1, 69117 Heidelberg, Germany}
\address{$^2$ Belarusian State University, 4 Nezavisimosty Ave., 220030, Minsk,   Belarus}

\begin{abstract}
The interaction Hamiltonian of an electron and a quasi-monochromatic pulse of a strong quantized electromagnetic field is examined. Canonical transformations of the field variables are found that allow the division of the system's Hamiltonian in two parts. The first one describes the interaction between an  electron and a single collective mode of the field. The properties  of this mode are defined by the superposition of the modes corresponding to the pulse wave packet. The second part describes the field fluctuations relatively to the collective mode. The field intensity, pulse duration and transversal spread are estimated for which a single-mode approximation can be used for the system's description.
\end{abstract}


\section{Introduction}\label{intro}
The   quantum electrodynamical (QED) processes in the presence of a strong electromagnetic field are commonly described in the Furry representation \cite{PhysRev.81.115} with scattering amplitudes calculated via the ``dressed'' external electron states  \cite{Nikishov1,Goldman,Ritus,Reiss,Ritus1985} instead of the free plane waves in vacuum QED. These states are defined by the exact solutions of the Dirac equation with an external electromagnetic field that can be  classical \cite{Volkov} or quantized \cite{Berson}. In the classical case, these exact solutions can be obtained when the external field is described by a monochromatic plane wave \cite{Volkov} or by a function which depends only on the field phase $\phi$, corresponding to the propagation of a free electromagnetic wave \cite{LandauQED}. In the quantum case, they are determined for a  quantized external field (single-mode approximation).

The Furry approach was successfully employed for the theoretical description of many QED processes, such as multi-photon (non-linear) Compton scattering \cite{PhysRevA.85.062102,Karlovets,Piazza,BocaOprea,Seipt,PhysRevLett.105.063903,BocaFlorescu,HarveyHeinzl,Gavrila,BergouVarro}, electron--positron pair creation \cite{PhysRevA.87.062107,PhysRevA.86.052104,Titov,Mueller,DiPiazza-Milstein}, sequential and non-sequential double ionization of atoms \cite{Becker1} and other problems  \cite{KeitelReview,Ehlotzky}. At the same time one should take into account that experimentally available laser pulses (external field) have a finite duration and a transversal spread, corresponding to a multi-mode wave packet with a non-zero spectral width and an angular divergence. As a consequence, the wave packet is not described by the exact solutions  \cite{Volkov}, \cite{Berson}. Therefore, it is of great importance to formulate the accurate conditions of the applicability of the single-mode approximation   in order to compare theoretical and experimental results.

An analogous problem in quantum optics is the analysis of the evolution of an atom interacting with a resonant field in a cavity. This system was initially described by Jaynes and Cummings \cite{JaynesCummings} who considered the interaction between an electromagnetic field and a two-level atom inside an ideal cavity. Although a lot of various aspects of the atom-field interaction were discussed through this model (for example, \cite{Knight} and references therein), in most works the external field was the single-mode cavity eigenstate. Few extensions to the model were proposed. One of them is based on the inclusion of the loses of the resonant mode through a lossy cavity \cite{PhysRevA.29.2627,PhysRevA.33.2444,PhysRevA.33.3610,PhysRevA.34.3077,PhysRevA.35.3433,PhysRevA.37.3175,PhysRevA.42.6884,PhysRevA.43.346,PhysRevA.44.4541,PhysRevA.44.6092,PhysRevA.47.2221}. Another one is the generalization for the case of few discrete modes interacting with an atom \cite{PhysRevA.40.5116,PhysRevA.42.4336,PhysRevA.42.6873,PhysRevA.44.6043,PhysRevA.45.6610}. However, models taking into account a superposition of modes in the vicinity of a resonant one have been discussed only recently \cite{PhysRevA.69.043807,PhysRevA.77.043822,PhysRevA.83.053826,PhysRevA.87.043817,PhysRevLett.110.100405,PhysRevLett.110.160401}.

In the present paper the mathematical background of the single-mode (plane wave) approximation for the description of a relativistic electron in a quantized field of a quasi-monochromatic laser pulse is investigated for the first time. The canonical transformation  of the field variables is found that allows one to separate the Hamiltonian of this system in two parts. The first term defines the interaction between an electron and a single collective mode. The parameters of this mode are related to the external field wave packet. The second part describes the field fluctuations with respect to the collective mode. Then the integral field intensity is determined for which these field fluctuations can be neglected, leading to the applicability conditions of the single-mode approximation.

The paper is organized as follows. In Sec. \ref{sec2}, the Hamiltonian of the system is considered and the qualitative characteristic parameters of the laser pulse are discussed. In Sec. \ref{sec3}, the canonical transformations of the field variables are performed and two the most relevant parts in the system's Hamiltonian are identified. In Sec. \ref{sec4}, the corrections to the single-mode Hamiltonian are discussed and the field intensity is determined for which the single-mode approximation is valid.

\section{Dirac equation for an electron in a multi-mode external field and characteristics of a laser pulse}\label{sec2}

Let us start with the equation for the state vector of the system which includes a relativistic electron and a multi-mode transversal quantized field $(\hbar = c =1)$

\begin{equation}\label{1}
i\frac{\partial \Psi}{\partial t} = \left(\sum_{\bi k} \om_{\bi k} a^\dag_{\bi k} a_{\bi k} + \boldsymbol \alpha\cdot(\bi p - e\bi A) + \beta m\right)\Psi,
\end{equation}
with the vector potential
\begin{equation}\label{2}
\bi A = \sum_{\bi k}\frac{\bi e (\bi k) }{\sqrt{2\omega_{\bi k} V}}\left(a_{\bi k} e^{i\thp{k}{r}}+a^\dag_{\bi k} e^{-i\thp{k}{r}}\right).
\end{equation}

Equation (\ref{1}) includes Dirac matrices $\boldsymbol \alpha$ and $\beta$,   a normalization volume $V$, a photon wave vector $ \bi k $, a frequency $\om_{\bi k}$ and a polarization vector $\bi e (\bi k)$, $\bi k \cdot \bi e(\bi k) = 0 $, photon annihilation and creation operators $a_{\bi k}$ and $a^\dag_{\bi k}$ of the mode $\bi k$, an electron charge $e$ and mass $m$. Equation (\ref{1}) can be written in covariant form if the transformation $\Psi = e^{-i \sum_{\bi k}\omega_{\bi k} t a^\dag_{\bi k} a_{\bi k}}\psi$ is used, thus obtaining the covariant form of the Dirac equation:
\begin{equation}\label{3}
	\left(i\gamma^\mu \partial_\mu - \gamma^\mu eA_\mu - m\right)\psi = 0,
\end{equation}
with the four product defined as $\fp{k}{x} = k^0 t - \thp{k}{x}, k^0 \equiv \om_{\bi k}$, $\partial_\mu = \partial/\partial x^\mu $, the metric tensor $g^{\mu\nu} = \mathrm{diag}(1,-1,-1,-1)$, the four potential of the field
\begin{equation*}
	A_\mu = \sum_{\bi k}\frac{e_{\mu}(k) }{\sqrt{2\omega_{\bi k} V}}\left(a_{\bi k} e^{-i\fp{k}{x}}+a^\dag_{\bi k} e^{i\fp{k}{x}}\right), \quad e^0 = 0,
\end{equation*}
and summations over repeated indices.
With the transformation $\psi = e^{i\sum_{\bi k} \fp{k}{x}a^\dag_{\bi k} a_{\bi k}}\chi$ the electron coordinates can be excluded from the field operators. As a result, the operators are transformed as follows:
\begin{eqnarray*}
i\gamma^\mu\partial_\mu  \rightarrow i\gamma^\mu\partial_\mu - \sum_{\bi k} \gamma^\mu k_\mu a^\dag_{\bi k} a_{\bi k},
\\
a_{\bi k} \rightarrow a_{\bi k} e^{i\fp{k}{x}},\quad a^\dag_{\bi k} \rightarrow a^\dag_{\bi k} e^{-i\fp{k}{x}},
\end{eqnarray*}
and equation (\ref{3}) leads to
\begin{equation}\label{4}
\left(i\hat\partial - \sum_{\bi k}\hat k a^\dag_{\bi k} a_{\bi k} - \sum_{\bi k}\hat b(\bi k) (a_{\bi k}+a^\dag_{\bi k}) - m\right)\chi = 0,
\end{equation}
\noindent where $b_{\mu} (\bi k) = e e_{\mu}(\bi k)/\sqrt{2V\om_{\bi k}}$, $e_{0}(\bi k) = 0$ and $\hat f\equiv \gamma^\mu f_\mu$ for any four-vector $f$. With the transformation $\chi = e^{-i\fp{q}{x}}\vp$, the coordinate dependence is separated from the field and spin degrees of freedom
\begin{equation}\label{5}
\left(\hat q - m - H \right)\vp = \left(\hat q - \sum_{\bi k}\hat k a^\dag_{\bi k} a_{\bi k} - \sum_{\bi k}\hat b(\bi k) (a_{\bi k}+a^\dag_{\bi k}) - m\right)\vp = 0,
\end{equation}
where the four vector $q$ can be considered as the total momentum of the system \cite{Berson}. The final form of equation (\ref{5}) will be used below.

Experimentally available laser pulses, can be described by a quasi-monochromatic wave packet with the central frequency $\omega_0$ and the wave vector $\bi k_0 = \omega_0 \bi n$ ($\bi n$ is a unit vector) with corresponding spreads  in a solid angle $\Delta \Omega$:
\begin{eqnarray}\label{6}
	\delta \omega \sim \frac{1}{\tau}, \quad \delta \bi k_0 \approx  \omega_0^2 \Delta \Omega \sim \frac{1}{S},
\end{eqnarray}
characterizing by the duration $\tau$ and spacial width $S$ of the laser pulse. The non-monochromaticity will be characterized via two dimensionless parameters
\begin{eqnarray}\label{7}
	\sigma_{2}= \frac{\delta \omega}{\omega_0} \approx \frac{1}{\omega_0 \tau},\quad	\sigma_1 = \frac{\delta \bi k_0 }{k_0^2} \approx \frac{1}{\omega_0^2 S}.
\end{eqnarray}

For high intensity pulses, as those considering in the following all modes within the volume $\Delta   = \delta \omega \delta \bi k_0$ in the $\bi k$-space are highly populated and correspond to the large quantum numbers $n_{\bi k}$ of the field state vector.

When only one mode is included in equation (\ref{5}), the single-mode approximation is recovered that leads to the Berson's solution \cite{Berson} in the case of a quantized field or to the Volkov's solution \cite{Volkov} in the case of a classical field. This intuitive conclusion will be thoroughly justified by using the Hamiltonian (\ref{5}) to consistently derive the single-mode approximation with correcting terms, appearing due to the interaction of the field modes between each other.

\section{Approximating single-mode Hamiltonian and canonical transformation for its diagonalization}\label{sec3}

In this section we employ the method of approximating Hamiltonian described in detail in reference \cite{Bogolubov_book}. Since the non vanishing modes of the quantized external field are inside a small volume $\Delta$ in $\bi k$-space the total Hamiltonian can be written as:
\begin{eqnarray}\label{8}
H_A &=&  \sum_{\bi k < \Delta}[ \hat k_0 a^\dag_{\bi k} a_{\bi k} +  \hat b_0 (a_{\bi k}+a^\dag_{\bi k})] + \sum_{\bi k > \Delta}\hat k a^\dag_{\bi k} a_{\bi k},
\\
H &\equiv& H_A  + H_1 + H_2. \nonumber
\end{eqnarray}
where constant four vectors $k_0$, $b_0$ and a small volume $\Delta$ in $k$-space near $k_0$ are the variational parameters of the approximating Hamiltonian and will be defined later. The sums $\sum_{\bi k < \Delta}$ and $\sum_{\bi k > \Delta}$ mean summation inside and outside the volume $\Delta$, respectively and the operators $H_{1,2}$ are determined identically from  equation (\ref{5})
\begin{eqnarray}
H_1 &=&  \sum_{\bi k < \Delta}\left[ (\hat k - \hat k_0) a^\dag_{\bi k} a_{\bi k} +  (\hat b(\bi k) - \hat b_0) (a_{\bi k}+a^\dag_{\bi k})\right], \label{9}
\\
H_2 &=& \sum_{\bi k > \Delta}\hat b(\bi k)(a_{\bi k}+a^\dag_{\bi k}). \label{10}
\end{eqnarray}

By the definition in ref. \cite{Bogolubov_book}, the approximating Hamiltonian $H_A$ should quantitatively describe the system, be accurately diagonalizable and the perturbations due to the operators $H_{1,2}$ need to be small. For the diagonalization of $H_A$ let us utilize the method of canonical transformations, which was introduced by Bogolubov and Tyablikov for the polaron problem in the strong field limit \cite{Bogolubov}. For this purpose we go back to the coordinate representation in (\ref{8}):
\begin{eqnarray}\label{11}
H_A  &=&  \frac{1}{2}\hat k_0 \sum_{\bi k<\Delta }    (p_{\bi k}^2 + q_{\bi k}^2) + \hat b_0 \sqrt{2}\sum_{\bi k<\Delta }   q_{\bi k} + \sum_{\bi k > \Delta}\hat k a^\dag_{\bi k} a_{\bi k},
\\
q_{\bi k} &=& \frac{a_{\bi k}+a^\dag_{\bi k}}{\sqrt{2}}, \quad p_{\bi k} = - i \frac{\partial}{\partial q_{\bi k} } = - i \frac{a_{\bi k} - a^\dag_{\bi k}}{\sqrt{2}}. \nonumber
\end{eqnarray}
Following Bogolubov \cite{Bogolubov}, we introduce the collective variable $Q$ in which all field modes are added coherently and the ``relative'' field variables $y_{\bi k}$ which define quantum fluctuations relative to the collective mode
\begin{equation}
	\label{12}
	\eqalign{Q   =  \sum_{\bi k<\Delta }   q_{\bi k}, \quad y_{\bi k} = q_{\bi k} - \frac{1}{N} Q,\quad q_{\bi k} = y_{\bi k} + \frac{1}{N} Q, 
	\cr
 	\sum_{\bi k<\Delta }y_{\bi k} = 0, \quad  N = \sum_{\bi k<\Delta}1,}
\end{equation}
where $N \gg 1$  is equal to the number of modes in the volume   $\Delta$. The transformation of the momentum operators is calculated according to its definition \cite{Bogolubov}:
\begin{eqnarray}\label{13}
  p_{\bi k}  =  - i \frac{\partial}{\partial q_{\bi k} }  = - i \left\{  \frac{\partial Q }{\partial q_{\bi k} }\frac{\partial}{\partial Q} + \sum_{\bi l<\Delta}  \frac{\partial y_{\bi l} }{\partial q_{\bi k} }\frac{\partial}{\partial y_{\bi l}}\right\}.
\end{eqnarray}
Calculation of the derivatives with the help of (\ref{12}) gives the generalized momenta:
\begin{equation}
	\label{14}
	\eqalign{p_{\bi k}  =  P + p_{y k}, \quad  \sum_{\Delta k } p_{y k} = 0, \quad P = - i \frac{\partial}{\partial Q}, 
	\cr
	p_{y k} =  - i \frac{\partial}{\partial y_{\bi k} } +   \frac{i}{N} \sum_{\Delta f }  \frac{\partial}{\partial y_f}.}
\end{equation}
Insertion of (\ref{12}) and (\ref{14}) into the Hamiltonian (\ref{11}) leads to the separation of the collective coordinates, the fluctuation operators $y_{\bi k}$ and the ``external'' variables $a_{\bi k}$ and $a^\dag_{\bi k}$, corresponding to $\bi k$ outside the $\Delta$ volume:
\begin{eqnarray}\label{15}
H_A  &=& \frac{1}{2}\hat k_0 \left[ \frac{1}{N}Q^2 + N P^2\right] + \hat b_0 \sqrt{2} Q \nonumber
\\
&+& \frac{1}{2}\hat k_0 \sum_{\bi k < \Delta  } (p_{yk}^2 + y_{\bi k}^2)  + \sum_{\bi k > \Delta}\hat k a^\dag_{\bi k} a_{\bi k}.
\end{eqnarray}

We can now quantize the collective and ``relative'' variables by introducing the new set of creation and annihilation operators
\begin{eqnarray}\label{16}
Q &=&  \frac{\sqrt{N} }{\sqrt{2}} (A + A^\dag), \quad P = -i \frac{1 }{\sqrt{2 N}} (A - A^\dag), \nonumber
\\
&&\big[A, A^\dag\big] = 1, \nonumber
\\
y_{\bi k}  &=& \frac{1}{\sqrt{2}}( \tilde{c}_{\bi k} + \tilde{c}_{\bi k}^\dag), \quad p_{yk}  = - i \frac{1}{\sqrt{2}}( \tilde{c}_{\bi k} -\tilde{ c}_{\bi k}^\dag),
\\
\tilde{c}_{\bi k} &=& a_{\bi k} - \frac{1}{N}\sum_{\bi l < \Delta  } a_{\bi l}, \quad [a_{\bi k}, a_{\bi k_1}^\dag] = \delta_{\bi k \bi k_1}, \nonumber
\\
&&\big[\tilde{c}_{\bi k} ,\tilde{c}_{\bi k_1}^\dag\big] = \delta_{\bi k \bi k_1} + \frac{1}{N}.\nonumber
\end{eqnarray}
With the accuracy of $ \sim 1/N$ the Hamiltonian (\ref{15}) transforms into
\begin{eqnarray}
H_A &=& \hat k_0 A^\dag A + \hat b_0 \sqrt{N}(A + A^\dag)\nonumber
\\
&&\mspace{46mu}+\hat k_0 \sum_{\bi k < \Delta  }   \tilde{c}_{\bi k}^\dag \tilde{c}_{\bi k} + \sum_{\bi k > \Delta}\hat k a^\dag_{\bi k} a_{\bi k}\label{17} 
\\
&&\mspace{46mu}\equiv H_{sm} + H_f + H_e,\nonumber
\\
H_{sm} &=& \hat k_0 A^\dag A + \hat b_0 \sqrt{N}(A + A^\dag), \label{18}
\\
H_f &=& \hat k_0 \sum_{\bi k < \Delta  }   \tilde{ c}_{\bi k}^\dag \tilde{c}_{\bi k}, \quad H_e = \sum_{\bi k > \Delta}\hat k a^\dag_{\bi k} a_{\bi k}. \label{19}
\end{eqnarray}
where the operators are written in the normal form and the energy of ``vacuum oscillations'' is not taken into account. In this representation the operator $H_{sm}$ which corresponds to the single-mode approximation is completely separated from the contributions defined by the fluctuation operator $H_f$ and by the external modes operator $H_e$. Therefore the state vector of the system in the zeroth approximation is represented as the product:
\begin{eqnarray}\label{20}
\eqalign{|\Psi^{(0)}\ra = |\Psi_A\ra |\{n_f\}\ra |\{n_e\}\ra ,  
\cr
\tilde{ c}_{\bi k}^\dag \tilde{c}_{\bi k} | n_{\bi k}^f\ra  =  n_{\bi k}^f |  n_{\bi k}^f \ra,
\cr
 a_{\bi k}^\dag a_{\bi k} | n_{\bi k}^e \ra  = n_{\bi k}^e| n_{\bi k}^e \ra,}
\end{eqnarray}
where $|n_{\bi k}^f \ra$ defines the state of the ``fluctuations'', $|n_{\bi k}^e \ra$ is the state of the ``external'' modes of the electromagnetic field, which does not interact with an electron, and $|\Psi_A\ra$ describes the state of the electron interacting with a collective mode of the field. $|\Psi_A\ra$ is determined as a solution of the equation:
\begin{eqnarray}
\Big\{ \hat  q^{(0)} - m - H_A\Big \}|\Psi_A\ra = 0  \label{21}
 \\
 H_A = \left\{\hat a + [\hat k_0 (A^\dag A + f) + \hat b_0 \sqrt{N}(A + A^\dag)]\right\}  |\Psi_A\ra,\label{22}
\\
a_{\mu} = \sum_{\bi k > \Delta}  k_{\mu}  n_{\bi k}^e, \quad f = \sum_{\bi k < \Delta}n_{\bi k}^f.\nonumber
\end{eqnarray}

The Hamiltonian $H_A$ up to the constant four vectors $a_{\mu}$ and $f_{\mu} = k_{0\mu} f$ coincides with the Dirac equation with the only one mode of the field and can be diagonilized. Solutions of equation (\ref{21}) in Bargmann representation for the creation and annihilation operators were found by Berson \cite{Berson} and their operator form were obtained in \cite{PhysRevA.87.052107}:
\begin{eqnarray}\label{23}
	|\Psi_{A}(n,a,f)\ra   = C_1\left(1+\frac{\hat k_0 \hat b_0 \sqrt{N}}{2\fp{z}{k_0}}(A+A^\dag)\right) Su(p_n)|n\rangle, 
	\\
	S=e^{\alpha(A^\dag - A)}e^{-\frac{\eta}{2}(A^2-A^{\dag2})}, \quad A^\dag A|n\rangle = n |n\rangle, \nonumber
	\\
	p_n = z-k_0\Bigg(\sqrt{1-\frac{2Nb_0^2}{\fp{z}{k_0}}}(n+1/2)-1/2 \nonumber
	\\
	\mspace{90mu}+f-\frac{\fp{z}{b_0}^2 N}{\fp{z}{k_0}^2}\frac{1}{1-2Nb_0^2/\fp{z}{k_0}}\Bigg), \nonumber
	\\
	\alpha = -\frac{\fp{z}{b_0}\sqrt{N}}{\fp{z}{k_0}}\frac{1}{1-2Nb_0^2/\fp{z}{k_0}},\nonumber
\\	
 \cosh \eta = \frac{1}{2}\left(\sqrt{\varkappa}+\frac{1}{\sqrt{\varkappa}}\right), \quad \varkappa = \frac{1}{\sqrt{1-\frac{2Nb_0^2}{\fp{z}{k_0}}}}, \nonumber
\end{eqnarray}
where $z_\mu = q_\mu^{(0)}- a_\mu$, $C_1$ is a normalization constant and $u(p_n)$ is a bispinor which coincides with a free Dirac bispinor.

In the general case parameters $z$ and $f$ in the quasi-momentum $p_n$ depend on the form of the laser pulse (see below equations (\ref{32}-\ref{38})). $p_n$ can be considered as the analog of the dressed momentum introduced in refs. \cite{PhysRevA.85.062102,PhysRevA.86.052104,PhysRevA.87.062107} and depends on the intensity and polarization $b_0$ of the collective mode. Let us note, that in the case of modulated plane waves, i. e., waves with few Fourier components the canonical transformation (\ref{12}) need to be modified for few collective modes and is not investigated in the present work.

The solutions (\ref{23}) form a full and orthogonal basis in the Hilbert space and can be normalized in a relativistically invariant way \cite{Ritus1985,Berson}.

\section{Perturbation theory on the operators $H_{1,2}$}\label{sec4}

It is supposed that the approximating eigenvectors (\ref{23}) define the main contribution to the solution of the initial equation (\ref{5}). In accordance with the method defined in ref. \cite{Bogolubov} one should now consider the corrections to this solution given by the operators $H_{1,2}$ from (\ref{9}) and (\ref{10}), thereby determining the optimal parameters ($\Delta, \hat k_0, \hat b_0$) of the approximating Hamiltonian (\ref{8}).

In order to build the corresponding perturbation theory let us insert a formal parameter $\lambda$ into equation (\ref{5})
\begin{eqnarray}\label{24}
	\left\{\hat  q  - m  - H_A\right\}  |\Psi\ra &=& \lambda (H_1+H_2)|\Psi\ra,
\end{eqnarray}
and represent a solution in a form of a series:
\begin{eqnarray}\label{25}
	|\Psi\ra = |\Psi^{(0)}\ra + \lambda |\Psi^{(1)}\ra + \ldots, \quad q = q^{(0)} + \lambda q^{(1)}  + \ldots.
\end{eqnarray}
From equation (\ref{25}) the first two orders of the perturbation can be found:
\begin{eqnarray}
(\hat q^{(0)} - m - H_A)|\Psi^{(0)}\ra = 0, \label{26}
	\\
\hat q^{(1)} |\Psi^{(0)}\ra + (\hat q^{(0)} - m - H_A)|\Psi^{(1)}\ra = (H_1+H_2)|\Psi^{(0)}\ra.\label{27}
\end{eqnarray}

Equation (\ref{26}) coincides with equation (\ref{21}) and has the set of eigenvectors (\ref{23}) which form a full and orthogonal basis in a Hilbert space. As was shown in \cite{Berson}, solution of equation (\ref{26}) leads to the Volkov's solution for an electron in a classical field \cite{Volkov} if one uses the coherent state representation instead of the Fock representation for the eigenvectors (\ref{23}). In order to do this in our case we suppose that the electromagnetic field is described by a set of coherent states:
\begin{eqnarray}\label{28}
|\Xi \ra = C \exp \left\{ \sum_{\bi k} [ u_{\bi k}a^\dag_{\bi k}  - u^*_{\bi k}a_{\bi k}]\right\} | 0 \ra, \quad  a_{\bi k}  | 0 \ra = 0,
\end{eqnarray}
with the amplitudes $u_{\bi k}$ associated with a field wave packet. This wave packet is localized in $\bi k$-space near the  momentum $\bi k_0$. The amplitudes $u_{\bi k}$ are modeled with the Gaussian distribution:
\begin{eqnarray}\label{29}
u_{\bi k} = e^{- \frac{\bi k^2_{\bot}}{2\sigma_1^2\omega_0^2}} e^{- \frac{(\omega - \omega_0)^2}{2\sigma_2^2\omega_0^2}}, \quad \bi k = \bi k_{\bot} + \omega \frac{\bi k_0}{\omega_0}, \quad \bi k_{\bot}\cdot\bi k_0 = 0,
\end{eqnarray}
where $\sigma_2$ and $\sigma_1$ are defined by equation (\ref{7}) and  determine a frequency and an angular spread in the laser pulse respectively. The Gaussian wave packet describes qualitative characteristics of the finite laser pulse and is convenient for the analytical calculations. The other choice of the wave packet form can change the obtained results on the number of the order of one.

The constant $C$ in (\ref{28}) for a pulse of intensity $I$,  transversal width $S$ and duration $\tau$ can be obtained from the normalization on the full pulse energy $W$:
\begin{eqnarray}
W &=& I S \tau = \langle \Xi|  \sum_{\bi k} \omega a^\dag_{\bi k} a_{\bi k} |\Xi \rangle \nonumber
\\
&&\mspace{29mu}= C^2 \frac{V}{8\pi^3}\int d\omega d \bi k_{\bot} \omega |u_{\bi k}|^2 \nonumber
\\
&&\mspace{29mu}= C^2 \frac{V}{ 8\pi^3 }\omega_0^4 \pi^{3/2}\sigma_1^2\sigma_2, \label{30}
\\
C &=& \sqrt{\frac{8\pi^{3/2}IS \tau}{V \sigma_1^2 \sigma_2 \omega_0^4}}.\nonumber
\end{eqnarray}

The state (\ref{28}) can be expanded in a series over the full set of states (\ref{23}):
\begin{eqnarray}\label{31}
|\Xi \ra = \sum_{n, a, f} C_n (a,f) |\Psi_{A}(n,a,f)\ra,
\end{eqnarray}
with coefficients $C_n (a,f)$, which depend not only on the collective mode quantum number $n$, but also on the ``fluctuating'' and the ``external'' modes quantum numbers $f$ and $a$ respectively. This linear combination can be used for the description of QED processes (non-linear Compton scattering, electron-positron pair creation) in Furry picture, taking into account the realistic  duration and angular spread of the laser pulse. In the following we show that the dependencies on $f$ and  $a$ can be neglected if the $\bi k$-space volume $\Delta$ is chosen in a consistent way.

We can estimate the contribution of the various terms in the Hamiltonian $H_A$ (\ref{22}) using the state (\ref{28}):
\begin{eqnarray}
	\label{32}
	\langle \Xi| \omega_0  A^\dag A|\Xi \rangle &\approx&  \langle \Xi|  \sum_{\bi k < \Delta} \omega_0 a^\dag_{\bi k} a_{\bi k} |\Xi \rangle \nonumber
	\\
	&=& \frac{\omega_0 C^2 V}{(2\pi)^3}\int_{\bi k < \Delta} d\bi k |u_{\bi k}|^2 = \nonumber
	\\
	&=& \frac{8 \pi^{3/2}IS \tau 2^3}{(2\pi)^3}\left(\int_0^{\frac{\Delta_1}{\sigma_1 \omega_0}}dt e^{-t^2}\right)^2 \int_0^{\frac{\Delta_2}{\sigma_2 \omega_0}}du e^{-u^2}\nonumber
	\\
	&=&  I S \tau \Phi^3(\delta), 
\end{eqnarray}
where we assumed that the volume $\Delta$ in $\bi k$-space can be written as $\Delta = \Delta_1^2 \Delta_2 = \delta^3 \sigma_1^2\sigma_2 \omega_0^3$. Here $\delta$ is a dimensionless parameter that will be defined below and $\Phi (z) = 2/\sqrt{\pi}\int_0^z e^{- t^2} dt$ is the error function. Other terms in $H_A$ (\ref{22}) are calculated in a similar way:
\begin{eqnarray}
\fl N =  \frac{V}{8\pi^3}\delta^3 \sigma_1^2\sigma_2 \omega_0^3, \label{33}
\\
\fl \langle \Xi|f|\Xi \rangle = \langle \Xi|\sum_{\bi k < \Delta}\tilde c^\dag_{\bi k}\tilde c_{\bi k}|\Xi \rangle =  \langle \Xi|  \sum_{\bi k < \Delta}  \left[a^\dag_{\bi k}- \frac{1}{N}\sum_{\bi l < \Delta}a^\dag_{\bi l}\right]  \left[a_{\bi k}- \frac{1}{N}\sum_{\bi l < \Delta}a_{\bi l}\right] |\Xi \rangle \nonumber
\\
= C^2\sum_{\bi k < \Delta}  \left[u^*_{\bi k}- \frac{1}{N}\sum_{\bi l < \Delta}u^*_{\bi l}\right]  \left[u_{\bi k}- \frac{1}{N}\sum_{\bi l < \Delta}u_{\bi l}\right] \nonumber
\\
=C^2 \frac{V}{(2\pi)^3}  \left(\int d \bi k |u_{\bi k}|^2 - \frac{V}{(2\pi)^3 N} \left| \int d \bi k  u_{\bi k} \right|^2\right)\nonumber
\\
 = IS \tau \Phi^3(\delta)\left(1-\frac{2^3 \pi^{\frac{3}{2}}}{\delta^3}\frac{\Phi^6(\frac{\delta}{\sqrt{2}})}{\Phi^3(\delta)}\right), \label{34}
\\
\fl \langle \Xi| a|\Xi \rangle \le \langle \Xi| a_0|\Xi \rangle =  \langle \Xi|  \sum_{\bi k >\Delta} \omega_0 a^\dag_{\bi k} a_{\bi k} |\Xi \rangle = I S \tau \left(1 - \Phi^3(\delta)\right),\label{35}
\end{eqnarray}

The contribution to the Hamiltonian $H_A$ due to the ``fluctuating'' modes is defined by the value $\langle \Xi|f|\Xi \rangle$ and is equal to zero if the parameter $\delta$ is chosen as the solution of the equation
\begin{eqnarray}\label{36}
1-\frac{2^3 \pi^{\frac{3}{2}}}{\delta^3}\frac{\Phi^6(\frac{\delta}{\sqrt{2}})}{\Phi^3(\delta)} = 0,\quad
\delta \approx 3.54.
\end{eqnarray}
It is evident that the actual value of this parameter depends on the laser pulse form but in any case it can be calculated in a similar way.

The contributions of the ``external'' pulse modes to $H_A$ can be neglected because they are defined by the value
\begin{eqnarray}\label{37}
\frac{\langle \Xi| a_0|\Xi \rangle}{\langle \Xi| \omega_0  A^\dag A|\Xi \rangle} = \frac{\left(1 - \Phi^3(\delta)\right)}{\Phi^3(\delta)}\approx 1.64\cdot 10^{-6},
\end{eqnarray}
\noindent when $\delta$ is found from (\ref{36}).

A similar estimation $\left(1 - \Phi^3(\delta)\right) \approx 10^{-6}$, defines the difference between the energy accumulated in the collective mode (\ref{32}) and the total energy of the laser pulse (\ref{30}). It means that if the parameter $\Delta$ is chosen as
\begin{eqnarray}\label{38}
\Delta \approx (3.54)^3 \sigma_1^2\sigma_2 \omega_0^3,
\end{eqnarray}
the values $a$ and $f$ can be omitted in the operator $H_A$ that corresponds to the vacuum of the ``fluctuating'' and ``external'' modes. In this case  $H_A$ can be considered as the single-mode Hamiltonian in the zeroth order (\ref{26}). The solution of this equation and its application for the analysis of the quantum corrections to the electromagnetic processes in the strong field were considered recently in our paper \cite{PhysRevA.87.052107}.

Now we determine the parameters $\hat k_0, \hat b_0, \omega_0$ of this Hamiltonian. Let us consider the first order equation (\ref{27}). Using its projection on the state vector $\la \Psi^{(0)}|$ (we pay attention to the fact, that for the correct perturbation theory for the Dirac equation the eigenvalue $\la \Psi^{(0)}|$ of a zero-order is not a hermitian conjugate to $| \Psi^{(0)}\ra$ but is the Dirac conjugate i. e. $\la \Psi^{(0)}| = \left(|\Psi^{(0)}\ra\right)^\dag \gamma^0$):
\begin{eqnarray}\label{39}
	 \la \Psi^{(0)}| \hat q^1  |\Psi^{(0)}\ra = \la \Psi^{(0)}|(H_1+H_2)|\Psi^{(0)}\ra.
\end{eqnarray}

According to Bogolubov \cite{Bogolubov} the stable solution of the initial equation (\ref{5}) exists if the first-order correction to the eigenvalue is equal to zero and this condition allows one to find the unknown parameters $\hat k_0, \hat b_0, \omega_0$. In our case it leads to two equations
\begin{eqnarray}\label{40}
 \la \Psi^{(0)}| H_1 |\Psi^{(0)}\ra = 0; \quad \la \Psi^{(0)}|H_2|\Psi^{(0)}\ra = 0,
\end{eqnarray}
because the operators $H_1$ and $H_2$ refer to different variables. The details of the calculations of the expectation values of the Hamiltonian $H_1$ can be found in appendix A. The result reads
\begin{align}\label{41}
		\la H_1 \ra &=  n_0\left(\vk+\frac{1}{\vk}\right)\left(\sum_{\bi k<\Delta}\frac{k}{N}-k_0\right)\cdot p_n \displaybreak[0]\nonumber
		\\
		&+ 2\frac{2 \alpha\vk n_0 + \alpha\left(\vk+\frac{1}{\vk}\right)n_0}{\sqrt{N}\fp{z}{k_0}}\left(\fp{b_0}{p_n}\sum_{\bi k<\Delta}\fp{k}{k_0}-\fp{k_0}{p_n}\sum_{\bi k<\Delta}\fp{k}{b_0}\right) \displaybreak[0]\nonumber
		\\
		&- \frac{n_0^2 (\vk+1/\vk)\vk+n_0^2(\vk-1/\vk)\frac{\vk}{2}}{\fp{z}{k_0}^2}\sum_{\bi k<\Delta} b_0^2 \fp{k}{k_0}\fp{k_0}{p_n}\displaybreak[0]\nonumber
		\\
		&+\frac{4\alpha}{\sqrt{N}}\left(\sum_{\bi k<\Delta}b(\bi k) - N b_0\right)\cdot p_n \displaybreak[0]\nonumber
		\\
		&+\frac{4 \vk n_0}{\fp{z}{k_0}}\Bigg(\left(Nb_0^2 - \sum_{\bi k<\Delta}\fp{b_0}{b(\bi k)}\right)\fp{k_0}{p_n}+ \fp{b_0}{p_n}\sum_{\bi k<\Delta}\fp{b(\bi k)}{k_0}\Bigg)\displaybreak[0]\nonumber
		\\
		&-\frac{12 \alpha\vk n_0\sqrt{N}}{\fp{z}{k_0}^2}b_0^2 \sum_{\bi k<\Delta}\fp{b(\bi k)}{k_0}\fp{k_0}{p_n}, 
\end{align}

The calculation of the average value of the Hamiltonian $H_2$ is performed in exactly the same way
\begin{eqnarray}\label{42}
	\fl\la H_2 \ra = \langle 0^f|\langle n_{\bi k}^e| \langle \Psi_A|H_2|\Psi_A\rangle|n_{\bi k}^e\rangle|0^f\rangle =\sum_{\bi k> \Delta}\langle \Psi_A|\hat b(\bi k)|\Psi_A\rangle\langle \Xi|(a_{\bi k}+a_{\bi k}^\dag)|\Xi\rangle \nonumber
	\\
	=\sum_{\bi k>\Delta}\Bigg(2 b(\bi k)\cdot p_n+ \frac{4\alpha \sqrt{N}}{\fp{z}{k_0}}\big(b(\bi k)\cdot k_0 b_0\cdot p_n\nonumber
	\\
	-b_0\cdot b(\bi k) k_0\cdot p_n\big)-\frac{2\vk n_0 N}{\fp{z}{k_0}^2}b_0^2 b(\bi k)\cdot k_0 k_0\cdot p_n\Bigg)\langle \Xi|(a_{\bi k}+a_{\bi k}^\dag)|\Xi\rangle.
\end{eqnarray}

As was stated above, according to ref. \cite{Bogolubov} the first corrections to the approximating Hamiltonian $H_A$ are equal to zero. This gives a condition for the determination of the variational parameters $\Delta$, $b_0$, $k_0$ of the Hamiltonian $H_A$. Therefore, if we choose 

\begin{eqnarray}\label{43}
\omega_0 =  \frac{1}{N}\sum_{\bi k<\Delta}\omega_{\bi k}, \ k_0 =  \frac{1}{N}\sum_{\bi k<\Delta}k, \  b_0 =  \frac{1}{N}\sum_{\bi k<\Delta}b(\bi k),
\end{eqnarray}
the expectation value of the Hamiltonian $H_1$ turns into zero. The average of the Hamiltonian $H_2$ vanishes according to symmetry consideration as it is not bilinear over polarization vectors $\sum_{\bi k>\Delta} b(\bi k)$. The physical meaning of this choice is that the collective single-mode corresponds to an  average over the modes of a quasi-monochromatic wave packet.

We have determined the variational parameters of the approximating Hamiltonian and consequently can proceed with the estimation of the field intensity for which the single-mode approximation is valid. For this purpose, the second-order correction
\begin{eqnarray*}
	E_0^{(2)} = - \sum_{E_{0}\neq E_{0 \alpha}} \frac{|\la E_{0}|H_1|E_{0 \alpha}\ra|^2}{E_{0 \alpha} - E_0},
\end{eqnarray*}
to the system's energy needs to be calculated. Here $H_1$ is the perturbation operator (\ref{A1}).

As can be seen from equation (\ref{A1}) only the last two terms contribute to $E_0^{(2)}$. The state $|E_{0 \alpha}\ra$ is the wave function
\begin{eqnarray}\label{44}
	|0_fn_0\ra = \frac{u(p)}{\sqrt{2 \epsilon}}S |n_0\ra|0_f\ra,
\end{eqnarray}
where $\epsilon$ is the electron's energy, $n_0$ is the number of quanta in the ``collective'' mode, $|0_f\ra$ is the state of the field fluctuations and for simplicity we neglected the term proportional to $\hat k \hat b$. Let us rewrite the Hamiltonian $H_1$ in the variables $A$, $A^\dag$ of the ``collective'' mode and $\tilde c_{\bi k}$, $\tilde c_{\bi k}^\dag$ of the fluctuations respectively
\begin{eqnarray}\label{45}
	\fl H_1 = (\hat k - \hat k_0)\left(-\frac{1}{2\sqrt{N}}(A-A^\dag)(\tilde c_{\bi k} -\tilde  c_{\bi k}^\dag)+(A+A^\dag)\frac{(\tilde c_{\bi k} + \tilde c_{\bi k}^\dag)}{2\sqrt{N}}\right) \nonumber
	\\
	+\sqrt{2}(\hat b(\bi k)-\hat b_0)\left(\frac{\tilde c_{\bi k} + c_{\bi k}^\dag}{\sqrt{2}}+\frac{A+A^\dag}{\sqrt{2N}}\right).
\end{eqnarray}

Now we can calculate the transition matrix element $\la 0_f n_0|H_1|n_f n\ra$:
\begin{eqnarray}
	\fl\la 0_f n_0|H_1|n_f n\ra = \frac{\bar u(p)(\hat k - \hat k_0)u(p)}{2 \epsilon\sqrt{N}}\la n_0|S^\dag A^\dag S|n\ra \delta_{1_f,n_f} \nonumber
	\\
	+\frac{\bar u(p)(\hat b(\bi k) - \hat k_0)u(p)}{2 \epsilon}\left[\delta_{1_f,n_f}\delta_{n_0,n}+\frac{1}{\sqrt{N}}\la n_0|S^\dag (A+A^\dag)S|n\ra \delta_{0_f,n_f}\right].\label{46}
\end{eqnarray}
The use of the transformation (\ref{A4}) of the creation and annihilation operators of the ``collective'' mode by the operator $S$ and the calculation of the averages in a spin space yields
\begin{eqnarray}\label{47}
		\fl\la 0_f n_0|H_1|n_f n\ra =\frac{(\fp{p}{k}-\fp{p}{k_0})\delta_{1_f,n_f}}{\epsilon \sqrt{N}}\Bigg[\frac{1}{2}\left(\sqrt{\vk}+\frac{1}{\sqrt{\vk}}\right)\sqrt{n_0}\delta_{n_0-1,n}\nonumber
		\\
		+\frac{1}{2}\left(\sqrt{\vk}-\frac{1}{\sqrt{\vk}}\right)\sqrt{n_0+1}\delta_{n_0+1,n}\Bigg] +\frac{\fp{b(\bi k)}{p}-\fp{b_0}{p}}{\epsilon}\Bigg[\delta_{1_f,n_f}\delta_{n_0,n} \nonumber
		\\
		+\frac{\sqrt{\vk}(\sqrt{n_0+1}\delta_{n_0+1,n}+\sqrt{n_0}\delta_{n_0-1,n})}{\sqrt{N}}\delta_{0_f,n_f}\Bigg].
\end{eqnarray}
Consequently we can write down the second-order correction to the system's energy
\begin{eqnarray}\label{48}
	E_0^{(2)}\approx \sum_{\bi k<\Delta}\frac{|\fp{v}{k}-\fp{v}{k_0}|^2}{ 4N}\left(\frac{\vk_+^2 n_0}{\omega_0-\omega_{\bi k}}-\frac{\vk_-^2 (n_0+1)}{\omega_{\bi k}+\omega_0}\right)\nonumber
	\\
	+\sum_{\bi k<\Delta}|\fp{v}{b(\bi k)}-\fp{v}{b_0}|^2\left(\frac{1}{\omega_{\bi k}}-\frac{\vk}{\omega_0 N}\right),
\end{eqnarray}
where $\vk_+ = \sqrt{\vk}+1/\sqrt{\vk}$, $\vk_- = \sqrt{\vk}-1/\sqrt{\vk}$, $v_\mu = p_\mu / \epsilon = (1,\bi v)$ and $\bi v$ is an electron's velocity: $\bi v = (v_\perp,0,v_z)$.

Equation (\ref{48}) has four terms but only two are important. The term inversely proportional to the frequency difference $\omega_0 - \omega_{\bi k}$ describes a resonance and defines the frequency renormalization and the lifetime of the collective mode. The term inversely proportional to $\omega_{\bi k}$ defines the fluctuations arising due to the interaction between an electron and an external field. The remaining two terms can be neglected as the second one is not a resonance and the fourth one is inversely proportional to the normalization volume $V$ ($N\sim V$ and $b\sim 1/\sqrt{V}$).

In order to perform a summation in $\bi k$-space we firstly fix a coordinate system. Let the $z$-axis be directed along $\bi k_0$, the $x$-axis along $\bi v_\perp$ - the velocity component perpendicular to the $\bi k_0$. The details of the summations over $\bi k$ can be found in appendix B. The second-order correction to the system's energy reads
\begin{eqnarray}\label{49}
	E_0^{(2)} &=& \frac{n_0 \omega_0 \pi v_\perp^2}{4}\left(\lambda \tan^{-1}4 \lambda - \frac{1}{4}+\frac{i \pi}{2}\cdot\left\{\begin{split}\lambda, \quad \lambda>1\\\frac{1}{\lambda}, \quad \lambda<1\end{split}\right.\right)\nonumber
	\\
	&+&e^2\frac{\delta ^5 \sigma _1^2 \sigma _2 \omega _0 \left(\sigma _2^2 \left(v^2-v_z^2\right){}^2+16 \sigma _1^2 v_\perp^2 v_z^2\right)}{96 (2\pi)^3 v^2},
\end{eqnarray}
where $\lambda = \sigma_2/(\delta \sigma_1^2)$.

The first term in (\ref{49}) is proportional to the same quantum number $n_0$ as the energy of the ``collective'' mode. Its real part defines the shift $\Delta \omega_0$ 
and the imaginary part defines the width $\Gamma/2$ of the resonant collective mode. One can rely on the single-mode approximation if these values are small
in comparison with $\omega_0$:
\begin{eqnarray}
	\frac{\Delta \omega_0}{\omega_0} = \frac{  \pi v_\perp^2}{4} \left[\lambda \tan^{-1}4 \lambda   - \frac{1}{4}\right] \ll 1, \label{50}
	\\
\frac{\Gamma}{\omega_0} =  \frac{  \pi^2 v_\perp^2}{4}\cdot  \left\{\begin{split}\lambda, \quad \lambda>1\\\frac{1}{\lambda}, \quad \lambda<1\end{split}\right.  \ \ll 1.\label{51}
\end{eqnarray}

As was stated above, the single-mode approximation is valid when the change in $E_0^{(2)}$ due to the fluctuations of the quantum field are small with comparison to the ground state energy of the ``collective'' mode (\ref{32}). These fluctuations are defined by the last term in equation (\ref{49}). This  leads to the additional parameter
\begin{eqnarray}\label{52}
	\mu =  e^2\frac{\delta ^5 \sigma _1^2 \sigma _2 \omega _0 \left(\sigma _2^2 \left(v^2-v_z^2\right){}^2+16 \sigma _1^2 v_\perp^2 v_z^2\right)}{96 (2\pi)^3 v^2 I S \tau \Phi^3(\delta)} \ll 1,
\end{eqnarray}
which defines the lowest pulse intensity for which the single-mode approximation can be used (see also \cite{PhysRevA.87.052107}) .

Modern lasers can reach nowadays high intensities \cite{PhelixFacility,Gerstner,Yanovsky,ELI,HIPER} up to $10^{22}\  \mathrm{W}/\mathrm{cm}^2$ with a pulse duration of about $30\, \mathrm{fs}$. Let us estimate the parameters (\ref{50}), (\ref{51}) and (\ref{52}) for an intensity $I = 10^{22}\  \textrm{W}/\textrm{cm}^2$,  photon frequency $\om = 7.8\cdot10^4 \ \textrm{cm}^{-1}$ (a corresponding wavelength of $800$ nm), pulse duration $\tau =  8.7\cdot10^{-4}\ \text{cm}^{-1}$ (corresponding to $30\ \text{fs}$) and focusing $S = 10^{-8}\ \text{cm}^2$.

The physical parameters $\sigma_1$ and $\sigma_2$ are connected with the characteristics of the laser pulse by equation (\ref{7}), and their numerical value for the above $I, S, \tau$ is equal to
\begin{eqnarray}
	\label{53}
	\sigma_1 = 0.127, \quad \sigma_2 = 0.014.
\end{eqnarray}

An electron beam always has angular divergence $\Delta \theta$ and $v_\perp \sim \Delta \theta\sim 1/\gamma$, where $\gamma$ is the electron's gamma factor. Therefore, for the moderately relativistic electrons we can consider that $v_\perp\leq \sigma_1$. 

By plugging the numerical values in equations (\ref{50}), (\ref{51}) and (\ref{52}) one obtains
\begin{eqnarray}
	\mu \sim 10^{-28}, \quad \frac{\Delta \omega}{\omega_0} \sim 6\cdot 10^{-4},\label{54}
	\\
	\frac{\Gamma}{\omega_0} \sim 0.01, \label{55}
\end{eqnarray}
and we can conclude that the single-mode approximation is applicable.

As can be seen from equation (\ref{54}) the parameter $\mu$ and frequency shift are very small values for the intensities in the strong field QED range, i. e., $10^{16}-10^{22}\ \text{W}/\text{cm}^2$ and pulse duration of 30 fs. This means that the electron mainly interacts with the collective single-mode. The influence of the fluctuations is suppressed. 

However, the most important parameter, which can limit the applicability of the single-mode approximation is indeed independent on the intensity, but depends on the pulse duration. It determines the width of the collective mode $\Gamma/\omega_0$, equation (\ref{55}). The decrease of the pulse duration from 30 fs to 3 fs, will increase its value by one order. The physical meaning of this result, corresponds to the situation that for the really short laser pulses, the collective field mode does not have sufficient time for its formation.

Concluding we can state, that the applicability of the single-mode approximation is mainly limited not by the pulse intensity, but rather by its duration and focusing size. Therefore, for a particular spectral distribution of the external laser pulse one should estimate $\sigma_1$ and $\sigma_2$, then insert their values together with the transversal electron velocity $v_\perp$ into equations (\ref{50}-\ref{52}) and make the conclusion about the applicability of the single-mode approximation. 

\section{Conclusion}
In this paper we have studied the applicability of the single-mode approximation for a relativistic electron interacting with a laser pulse of a finite duration and transversal width. The relations between parameters of the single-mode Hamiltonian and the form of the wave packet are found. In particular, the frequency of the collective mode corresponds to the average frequency of the wave packet's modes. 

The three parameters which determine the applicability of the single-mode approximation are frequency shift and width of the collective mode, and dimensional parameter $\mu$, which is defined as the relation of the energy of the fluctuations to the ground state energy of the collective mode. These parameters are determined by the physical parameters of the laser pulse, namely the field intensity, pulse duration and focusing size. For sufficiently long laser pulses and for experimentally available intensities the single-mode approximation is proven to be valid, however for the very short laser pulses the collective field mode does not have sufficient time to form. 

The proposed approach can also be used for the analysis of the interaction between an atom and a field placed in a non-ideal cavity.

\ack
The authors are grateful to C. H. Keitel, K. Hatsagortsyan and S. Cavaletto for fruitful discussions. OS acknowledges financial support from Max Planck Institute for Nuclear Physics.

\appendix
\section*{Appendix A\label{appendixA}}
\setcounter{section}{1}

We start with the calculation of the average of the Hamiltonian $H_1$. For this purpose we need to rewrite it in the variables of a ``collective'' mode $A$, $A^\dag$ and ``fluctuations'' $\tilde c_{\bi k}$, $\tilde c_{\bi k}^\dag$.
\begin{align}\label{A1}
	H_1 &= \sum_{\bi k< \Delta} (\hat k - \hat k_0)a_{\bi k}^\dag a_{\bi k} + (\hat b(\bi k) - \hat b_0)(a_{\bi k}+a_{\bi k}^\dag)\displaybreak[0] \nonumber
	\\
	&= \sum_{\bi k< \Delta}(\hat k - \hat k_0)\frac{1}{2}(p_{\bi k}^2 + q_{\bi k}^2)+\sqrt{2} (\hat b(\bi k) - \hat b_0)q_{\bi k} \displaybreak[0] \nonumber
	\\
	&= \sum_{\bi k< \Delta} (\hat k - \hat k_0)\frac{1}{2}\left(P^2 + 2P p_{yk}+ p_{yk}^2 + y_{\bi k}^2 + \frac{2Q y_{\bi k}}{N}+\frac{Q^2}{N^2}\right)\nonumber
	\\
	&\mspace{90mu}+ \sum_{\bi k< \Delta}\sqrt{2}(\hat b(\bi k) - \hat b_0)\left(y_{\bi k} + \frac{1}{N}Q\right)  \displaybreak[0]\nonumber
	\\
	&=\sum_{\bi k< \Delta}\frac{1}{2}(\hat k - \hat k_0)\left(P^2+\frac{Q^2}{N^2}\right) + \sum_{\bi k< \Delta}\frac{1}{2}(\hat k- \hat k_0)(p_{yk}^2 + y_{\bi k}^2) 
	\\
	&\mspace{90mu}+\sum_{\bi k< \Delta}(\hat k - \hat k_0)\left(P p_{yk}+ Q \frac{y_{\bi k}}{N}\right)+ \sum_{\bi k< \Delta}\sqrt{2}(\hat b(\bi k) - \hat b_0)\left(y_{\bi k} + \frac{Q}{N}\right).\nonumber
\end{align}
The expectation value of $H_1$, computed with respect to the ground state of fluctuations (\ref{17}) is equal to
\begin{eqnarray}
	\fl\la H_1\ra = \langle 0^f|\langle n_{\bi k}^e| \langle \Psi_A|H_1|\Psi_A\rangle|n_{\bi k}^e\rangle|0^f\rangle \nonumber
	\\
	= \langle \Psi_A|A^\dag A\left(\sum_{\bi k< \Delta}\frac{\hat k}{N}-\hat k_0\right)+\frac{(A^\dag + A)}{\sqrt{N}}\left(\sum_{\bi k< \Delta}\hat b(\bi k) - N\hat b_0\right)|\Psi_A\rangle.\label{A2}
\end{eqnarray}
By exploiting the definition of $|\Psi_A\ra$ in equation (\ref{A2}), one obtains
\begin{eqnarray}\label{A3}
	\fl\la H_1\ra = \langle n| \bar u(p_n) S^\dag \Bigg(A^\dag A  \left(\sum_{\bi k} \frac{\hat k}{N}-\hat k_0\right)\nonumber
	\\
\fl	+\frac{\hat b_0 \hat k_0 \sum_{\bi k} \hat k}{2\sqrt{N}\fp{z}{k_0}}(A^\dag+A)A^\dag A+A^\dag A (A+A^\dag)\frac{\sum_{\bi k} \hat k \hat k_0 \hat b_0}{2\sqrt{N}\fp{z}{k_0}}- \nonumber
	\\
	\fl-\frac{\sqrt{N}(A+A^\dag)A^\dag A}{2\fp{z}{k_0}} \hat b_0 \hat k_0 \hat k_0+\frac{\hat b_0 \hat k_0 \sum_{\bi k} \hat k \hat k_0 \hat b_0}{4\fp{z}{k_0}^2}(A+A^\dag)A^\dag A(A+A^\dag)\Bigg)S u(p_n)|n\rangle+ \nonumber
	\\
	\fl+\langle n|\bar u(p_n)S^\dag\Bigg(\frac{(A+A^\dag)}{\sqrt{N}} \left(\sum_{\bi k} \hat b(\bi k) - N\hat b_0\right)+\frac{\hat b_0 \hat k_0 (\sum_{\bi k} \hat b(\bi k) - N \hat b_0)}{2\fp{z}{k_0}}(A+A^{\dag})^2+ \nonumber
	\\
	\fl+\frac{(\sum_{\bi k} \hat b(\bi k) - N \hat b_0)\hat k_0 \hat b_0 }{2\fp{z}{k_0}}(A+A^{\dag})^2\nonumber
	\\
	\fl+\frac{\sqrt{N}\hat b_0 \hat k_0 (\sum_{\bi k} \hat b(\bi k) - N\hat b_0)\hat k_0 \hat b_0}{4\fp{z}{k_0}^2}(A+A^\dag)^3\Bigg)S u(p_n)|n\rangle,
\end{eqnarray}
where $S = e^{\alpha(A^\dag - A)}e^{-\frac{\eta}{2}(A^2-A^{\dag2})}$. Next we calculate the average of the field variables, taking into account the transformation law of $A$ and $A^\dag$ by operator $S$:
\begin{eqnarray}\label{A4}
	S^\dag A S =  \frac{1}{2}(\sqrt{\vk}+\frac{1}{\sqrt{\vk}})A+\frac{1}{2}(\sqrt{\vk}-\frac{1}{\sqrt{\vk}})A^\dag + \alpha, \nonumber
	\\
	S^\dag A^\dag S = \frac{1}{2}(\sqrt{\vk}+\frac{1}{\sqrt{\vk}})A^\dag+\frac{1}{2}(\sqrt{\vk}-\frac{1}{\sqrt{\vk}})A + \alpha.
\end{eqnarray}
Parameter $\vk$ was defined in equation (\ref{23}). Therefore, we can find how the combination of $A$ and $A^\dag$ in equation (\ref{A3}) transforms, for example:
\begin{eqnarray}\label{A5}
	\fl A^\dag A \rightarrow \frac{1}{2} \left(\vk+\frac{1}{\vk}\right) + \frac{1}{4}\left(\vk-\frac{1}{\vk}\right)\left(A^2+A^{\dag2}\right)+\alpha\sqrt{\vk}\left(A+A^\dag\right)+\beta,  
\end{eqnarray}
where $\beta = \left(\alpha^2 +\frac{1}{4}\left(\sqrt{\vk}-\frac{1}{\sqrt{\vk}}\right)^2\right)$.

According to the definition, the collective mode has a high intensity, i. e. it is highly populated, with the quantum number $n = n_0$ being a large value. Therefore, the averages with respect to the field variables are
\begin{eqnarray}\label{A6}
	\fl\la n_0|S^\dag (A+A^\dag) S|n_0\ra = 2 \alpha, \quad	\la n_0|S^\dag (A+A^\dag)^2 S|n_0\ra = \vk(2n_0+1) + 4 \alpha^2, \nonumber
	\\
	\fl\la n_0|S^\dag (A+A^\dag)^3 S|n_0\ra = 6 \alpha \vk (2n_0+1)+8 \alpha^3, \quad \langle n_0|S^\dag A^\dag A S|n_0\rangle = \frac{n_0}{2}\left(\vk+\frac{1}{\vk}\right)+\beta, \nonumber
	\\
	\fl\la n_0|S^\dag (A+A^\dag)A^\dag A S|n_0\ra = \la n_0|S^\dag A^\dag A(A+A^\dag) S|n_0\ra \nonumber
	\\
	= \alpha\vk(2n_0+1)+\alpha\left(\vk+\frac{1}{\vk}\right)n_0+2 \alpha \beta, \nonumber
	\\
	\fl\la n_0|S^\dag (A+A^\dag)A^\dag A(A+A^\dag) S|n_0\ra = \frac{\vk}{2}\left(\vk+\frac{1}{\vk}\right)(2n_0^2 + n_0 +1)\nonumber
	\\
	+\frac{\vk}{4}\left(\vk-\frac{1}{\vk}\right)(2n_0^2+2n_0)
	+(\beta \vk+4 \alpha \vk)(2n_0+1)\nonumber
	\\
	+2 \alpha^2 \left(\vk+\frac{1}{\vk}\right)n_0 + 4 \alpha^2 \beta.
\end{eqnarray}
The last step is to calculate the averages in the Dirac spin space. For this purpose we will employ the electron's density matrix $u(p)\otimes \bar u(p) = \rho = 1/2(\hat p+ m)(1-\gamma^5 \hat a)$, with $p$ and $a$ being its four momentum and polarization, respectively. For example,
\begin{eqnarray}\label{A7}
	\fl u_{\beta}(p_n)\bar u_\alpha(p_n)\left(\frac{1}{N}\sum_{\bi k<\Delta}\hat k - \hat k_0\right)_{\alpha \beta} = \rho_{\beta \alpha} \left(\frac{1}{N}\sum_{\bi k<\Delta}\hat k - \hat k_0\right)_{\alpha \beta}= \textrm{Sp} \left(\rho \left(\frac{1}{N}\sum_{\bi k<\Delta}\hat k - \hat k_0\right)\right).
\end{eqnarray}
  
Inserting (\ref{A6}) and (\ref{A7}) into (\ref{A3}) we find the average value of the Hamiltonian
\begin{eqnarray}\label{A8}
		\fl\la H_1 \ra =  n_0\left(\vk+\frac{1}{\vk}\right)\left(\sum_{\bi k<\Delta}\frac{k}{N}-k_0\right)\cdot p_n \nonumber
		\\
		+ 2\frac{2 \alpha\vk n_0 + \alpha\left(\vk+\frac{1}{\vk}\right)n_0}{\sqrt{N}\fp{z}{k_0}}\left(\fp{b_0}{p_n}\sum_{\bi k<\Delta}\fp{k}{k_0}-\fp{k_0}{p_n}\sum_{\bi k<\Delta}\fp{k}{b_0}\right) \nonumber
		\\
		- \frac{n_0^2 (\vk+1/\vk)\vk+n_0^2(\vk-1/\vk)\frac{\vk}{2}}{\fp{z}{k_0}^2}\sum_{\bi k<\Delta} b_0^2 \fp{k}{k_0}\fp{k_0}{p_n}
		\\
		+\frac{4\alpha}{\sqrt{N}}\left(\sum_{\bi k<\Delta}b(\bi k) - N b_0\right)\cdot p_n \nonumber
		\\
		+\frac{4 \vk n_0}{\fp{z}{k_0}}\Bigg(\left(Nb_0^2 - \sum_{\bi k<\Delta}\fp{b_0}{b(\bi k)}\right)\fp{k_0}{p_n}+ \fp{b_0}{p_n}\sum_{\bi k<\Delta}\fp{b(\bi k)}{k_0}\Bigg)\nonumber
		\\
		-\frac{12 \alpha\vk n_0\sqrt{N}}{\fp{z}{k_0}^2}b_0^2 \sum_{\bi k<\Delta}\fp{b(\bi k)}{k_0}\fp{k_0}{p_n}, \nonumber
\end{eqnarray}
where only the leading terms in $n_0$ are left.

\appendix
\section*{Appendix B\label{appendixB}}
\setcounter{section}{2}

Then the first term in equation (\ref{48}) is
\begin{eqnarray}\label{B1}
	\fl\sum_{\bi k<\Delta}\frac{|(\omega_{\bi k}-\omega_0)-(\bi k - \bi k_0)\cdot \bi v|^2}{\omega_0 - \omega_{\bi k}} =\frac{V}{(2\pi)^3}\int_{\Delta}d \bi k\Bigg((\omega_0 - \omega_{\bi k})\nonumber
	\\
	+2 (\bi k - \bi k_0)\cdot \bi v+\frac{|(\bi k - \bi k_0)\cdot \bi v|^2}{\omega_0 - \omega_{\bi k}}\Bigg) = \frac{V}{(2\pi)^3}\int_{\Delta}d \bi k \frac{|(\bi k - \bi k_0)\cdot \bi v|^2}{\omega_0 - \omega_{\bi k}}
\end{eqnarray}
The first two integrals in equation (\ref{B1}) are equal to zero as integrals of an odd function over a symmetric interval. Now, we change a variable $\bi k$ to $\bi k_0 + \bi q$. After expansion of the scalar product in (\ref{B1}), the terms linear in $\bi q$ will also not contribute as they are the odd functions. For this reason, one obtains
\begin{eqnarray}\label{B2}
	\frac{V}{(2\pi)^3}\int_{\Delta}d \bi k \frac{|(\bi k - \bi k_0)\cdot \bi v|^2}{\omega_0 - \omega_{\bi k}} = \frac{V}{(2\pi)^3}\int_{\Delta}d \bi q \frac{(\thp{q}{v})^2(\omega_0 + \omega_{\bi k})}{\omega_0^2 - \omega_{\bi k}^2} \nonumber
	\\
	= \frac{2 \omega_0 V}{(2\pi)^3}\int_{\Delta}d \bi q \frac{(\thp{q}{v})^2}{\omega_0^2 - \omega_{\bi k}^2} = -\frac{2 \omega_0 V}{(2\pi)^3}\int_{\Delta}d \bi q \frac{v_\perp^2 q_x^2+v_z^2 q_z^2}{2 \omega_0 q_z + q_z^2+q_x^2+q_y^2}.
\end{eqnarray}
The denominator of equation (\ref{B2}) is equal to zero for $q_{z,1} = - 2\omega_0$, $q_{z,2} = -(q_x^2+q_y^2)/2\omega_0$ and needs to be regularized via \cite{Akhiezer}
\begin{eqnarray}\label{B3}
	\frac{1}{u} = P\frac{1}{u} - i \pi \delta(u),
\end{eqnarray}
with the symbol $P$ being the principal value. Insertion of (\ref{B3}) into (\ref{B2}) gives
\begin{eqnarray}\label{B4}
	 -\frac{2 \omega_0 V}{(2\pi)^3}\int_{\Delta}d \bi q \frac{v_\perp^2 q_x^2+v_z^2 q_z^2}{(q_z-q_{z,1})(q_z-q_{z,2})} \nonumber
	 \\
	 +  \frac{2 i \pi V \omega_0}{(2\pi)^3}\int d\bi q (v_\perp^2 q_x^2 + v_z^2 q_z^2)\delta((q_z-q_{z,1})(q_z-q_{z,2})) \nonumber
	 \\
	 =-\frac{V v_\perp^2}{(2\pi)^3}\int_{\Delta}d \bi q \frac{ q_x^2}{q_z+\frac{q_x^2+q_y^2}{2 \omega_0}} + \frac{i \pi V v_\perp^2}{(2\pi)^3}\int d\bi q  q_x^2 \delta\left(q_z+\frac{q_x^2+q_y^2}{2 \omega_0}\right).
\end{eqnarray}
While obtaining (\ref{B4}) we took into account that $(q_z-q_{z,1})(q_z-q_{z,2}) =2 \omega_0(q_z-q_{z,2})$ and the terms proportional to $v_z^2$ can be neglected, as the integration takes place near zero, leading to $q_z^2 \sim q_\perp^4$. The integration in the first term is performed in a polar coordinate system, yielding:
\begin{eqnarray}\label{B5}
	-\frac{V v_\perp^2}{(2\pi)^3}\int_{\Delta}d \bi q \frac{ q_x^2}{q_z+\frac{q_x^2+q_y^2}{2 \omega_0}} = -\frac{V v_\perp^2}{(2\pi)^3}\int d q_z d q_\perp d \phi \frac{q_\perp^3 \cos^2{\phi} }{q_z+\frac{q_\perp^2}{2 \omega_0}} \nonumber
	\\
	= -\frac{\pi V v_\perp^2}{(2\pi)^3}\left(\frac{\Delta_1^2 \Delta_2 \omega_0}{8} - \frac{\Delta_2^2 \omega_0^2}{2}\tan^{-1}\frac{4 \Delta_2 \omega_0}{\Delta_1^2}\right)
\end{eqnarray}

The integration of the second term with respect to $q_z$ yields a Heaviside function $\theta(\omega_0 \Delta_2 - q_{x}^2-q_{y}^2)$ as $q_z=-(q_{x}^2+q_{y}^2)/2 \omega_0> -\Delta_2/2$. Therefore,
\begin{eqnarray}\label{B6}
	  \frac{i \pi V v_\perp^2}{(2\pi)^3}\int d\bi q  q_x^2 \delta\left(q_z+\frac{q_x^2+q_y^2}{2 \omega_0}\right) \nonumber
	\\
	=   \frac{i \pi^2 V v_\perp^2}{(2\pi)^3}\cdot \left\{\begin{split}&\frac{(\omega_0 \Delta_2)^2}{4},\quad \sqrt{\omega_0 \Delta_2}<\Delta_1 \\ &\frac{\Delta_1^4}{4},\quad \sqrt{\omega_0 \Delta_2}>\Delta_1\end{split}\right.,
\end{eqnarray}
Combining together (\ref{B5}) and (\ref{B6}) one obtains
\begin{eqnarray}\label{B7}
	\fl\frac{V}{(2\pi)^3}\int_{\Delta}d \bi k \frac{|(\bi k - \bi k_0)\cdot \bi v|^2}{\omega_0 - \omega_{\bi k}} = \nonumber
	\\
	 \frac{\pi V v_\perp^2 \omega_0^4 \delta^2}{(2\pi)^3}\left(\frac{\delta \sigma_1^2 \sigma_2 }{8} - \frac{\sigma_2^2 }{2}\tan^{-1}\frac{4 \sigma_2 }{\delta \sigma_1^2}\right)\nonumber
	\\
	 + \frac{i \pi^2 V v_\perp^2}{(2\pi)^3}\frac{\omega_0^4}{4}\cdot \left\{\begin{split}&\delta^2 \sigma_2^2,\quad \sqrt{\frac{\sigma_2}{\delta}}<\sigma_1 \\ &\delta^4 \sigma_1^4,\quad \sqrt{\frac{\sigma_2}{\delta}}>\sigma_1\end{split}\right..
\end{eqnarray}

At last, we come to the calculation of the remaining correction in equation (\ref{48}) from the fluctuations:
\begin{eqnarray}\label{B8}
	\sum_{\bi k<\Delta}\frac{|\bi v (\bi b(\bi k)-\bi b_0)|^2}{\omega_{\bi k}} = \frac{e^2}{2V}\sum_{\bi k<\Delta,\alpha}\frac{|\bi v\cdot(\bi e_{k, \alpha}/\sqrt{\omega_{\bi k}}-\bi e_{0,\alpha}/\sqrt{\omega_0})|^2}{\omega_{\bi k}}\bi.
\end{eqnarray}
In order to calculate the integral in equation (\ref{B8}), we introduce the polarization vectors
\begin{eqnarray}\label{B9}
	\fl\bi e_{0,2} = \frac{\bi k_0\times \bi v}{\omega_0 v}, \quad \bi e_{0,1} = \frac{\bi e_{0,2}\times \bi k_0}{\omega_0} = \frac{(\bi k_0 \times \bi v)\times \bi k_0}{\omega_0^2 v} = \frac{\bi v \omega_0^2 - \bi k_0(\thp{k_0}{v})}{\omega_0^2 v}, \nonumber
	\\
	\fl\bi e_{k,2} = \frac{\bi k\times \bi v}{\omega_{\bi k} v}, \quad \bi e_{k,1} = \frac{\bi e_{k,2}\times \bi k}{\omega_{\bi k}} = \frac{(\bi k \times \bi v)\times \bi k}{\omega^2_{\bi k} v}=\frac{\bi v \omega^2_{\bi k} - \bi k(\thp{k}{v})}{\omega^2_{\bi k} v}.
\end{eqnarray}
Here we pay attention to the fact that $\bi v \cdot \bi e_{k,2}=\bi v \cdot \bi e_{0,2}=0$. Insertion of equation (\ref{B9}) into equation (\ref{B8}) gives
\begin{eqnarray}\label{B10}
	\frac{e^2}{2(2\pi)^3 \omega_0}\int d\bi k \left|v\left(\frac{1}{\sqrt{\omega_{\bi k}}}-\frac{1}{\sqrt{\omega_0}}\right) - \frac{1}{v}\left(\frac{(\thp{k}{v})^2}{\omega^{5/2}_{\bi k}}-\frac{(\thp{k_0}{v})^2}{\omega_0^{5/2}}\right)\right|^2.
\end{eqnarray}
The change of variable $\bi k $ to $\bi k_0 + \bi q$ and the decomposition of the differences in brackets up to the first order in Taylor series in $\bi q$ bring us to the final result
\begin{eqnarray}\label{B11}
	\frac{e^2}{2(2\pi)^3 \omega_0}\int d\bi q \left|\left(-\frac{v}{2 \omega_0^{5/2}}+\frac{5(\thp{k_0}{v})^2}{2 v \omega_0^{9/2}}\right)\thp{k_0}{q}-\frac{2\thp{k_0}{v}}{v\omega_0^{5/2}}\thp{v}{q}\right|^2 \nonumber
	\\
	= e^2\frac{\delta ^5 \sigma _1^2 \sigma _2 \omega _0 \left(\sigma _2^2 \left(v^2-v_z^2\right){}^2+16 \sigma _1^2 v_\perp^2 v_z^2\right)}{96 (2\pi)^3 v^2}.
\end{eqnarray}

\section*{References}
\bibliographystyle{unsrt}
\bibliography{quantum-2}

\end{document}